%% file: ETRIJ-v2.tex
\documentclass{article}

\usepackage{arxiv}
\usepackage[utf8]{inputenc} %
\usepackage[T1]{fontenc}    %
\usepackage{hyperref}       %
\usepackage{url}            %
\usepackage{booktabs}       %
\usepackage{amsfonts}       %
\usepackage{nicefrac}       %
\usepackage{microtype}      %
\usepackage{lipsum}		%
\usepackage{graphicx}
\usepackage{natbib}
\usepackage{doi}

\usepackage[monochrome]{xcolor}
\usepackage{algorithm, algorithmicx, algpseudocode}
\usepackage{graphicx}
\DeclareGraphicsExtensions{.pdf,.jpeg,.png}

\usepackage[labelformat=simple]{subfig} 

\usepackage{amssymb}
\usepackage{amsmath,graphicx}
\usepackage{amsfonts}
\usepackage{makecell}
\usepackage {multirow}
\usepackage {multicol}
\usepackage {enumitem}

\DeclareMathOperator*{\argmax}{argmax}

\usepackage[switch,pagewise]{lineno}

\newcommand\BLUE{black}

\begin{document}

\title{Fast offline Transformer-based end-to-end automatic speech recognition for real-world applications\protect}

\author{ {\hspace{1mm}Yoo Rhee Oh, Kiyoung Park \thanks{*Corresponding author: Kiyoung Park, \texttt{pkyoung@etri.re.kr}}, Jeon Gue Park}\\
	Artificial Intelligence Research Laboratory\\
	Electronics and Telecommunications Research Institute (ETRI)\\
	Daejeon, Republic of Korea \\
}

\renewcommand{\shorttitle}{Fast offline Transformer-based end-to-end ASR for real-world applications}

\maketitle
\input{abstract}

\input{intro}

\input{review}

\section{Fast and efficient Transformer-based speech recognition}
\label{sec:transformer2}

\input{batch}

\input{ctc}

\input{segment}

\section{Experiments}
\label{sec:expr}

\input{baseline}

\input{exp_intro}

\input{exp_batch}

\input{exp_segment}

\input{conc}

\section{Acknowledgement}
This work was supported by Institute for Information \& communications
Technology Planning \& Evaluation(IITP) grant funded by the Korea
government(MSIT) (No.2019-0-01376, Development of the multi-speaker
conversational speech recognition technology)

\bibliographystyle{unsrtnat}
\bibliography{total.bib}

\end{document}

%% file: abstract.tex
\begin{abstract}
		With the recent advances in technology, automatic speech recognition
		(ASR) has been widely used in real-world applications. The efficiency
		of converting large amounts of speech into text accurately
		with limited resources has become more important than ever.
		This paper proposes a method to rapidly
		recognize a large speech database via a Transformer-based end-to-end
		model. Transformers have improved the state-of-the-art performance in many
		fields. However, they are not easy to use for long
		sequences. In this paper, various techniques to speed up the recognition of
		real-world speeches are proposed and tested, including decoding
		via multiple-utterance batched beam
		search, detecting end-of-speech based on a connectionist temporal
		classification (CTC), restricting the CTC prefix score,
		and splitting long speeches into short segments.
		Experiments are conducted with the Librispeech English and
		the real-world Korean ASR tasks to verify the proposed methods.
		From the experiments, the proposed system can convert 8 hours of
		speeches spoken at real-world meetings into text in less than 3 minutes
		with a 10.73\% character error rate, which is 27.1\% relatively lower than
        that of conventional systems.
\end{abstract}

\keywords{Speech recognition, Transformer, end-to-end, connectionist temporal classification}

%% file: intro.tex
\section{Introduction}
\label{sec:intro}

Owing to the rapid advances in automatic speech recognition (ASR)
research~\cite{asr_review,asr_review2,asr_review3,asr_dnn}, these technologies have
been widely adopted in many practical applications, such as automatic response
systems, automatic subtitle generation, and meeting transcription.  {\color{\BLUE}Deep neural network (DNN)}
incorporated with the traditional hidden Markov Model (HMM) has worked
powerfully and raised speech recognition accuracy to the commercial level.

A recent breakthrough in deep learning from
natural language processing solved sequence generation
problem in a novel way~\cite{Sutskever14,Bahdanau2015}.  Further, a Transformer
model in which sequences are processed more thoroughly with self-attention
achieves state-of-the-art performances in many tasks.

In ASR, where the speech recognition problem is formulated as generation of the
most probable sequence of words given a spoken utterance as an input sequence,
a Transformer-based end-to-end speech recognition technique has achieved state-of-the-art
performance and a major breakthrough since deep learning technology was
used with HMM~\cite{transformer,speech_transformer}.

However, it is well-known that the Transformer is not sufficient at processing
long sequences with regard to performance and computation speed, and
speech signals are usually represented by much longer sequences than the
written sentence. There have been many research studies attempting to overcome this drawback,
including~\cite{Moritz19,Moritz20,transformer_mono}.

There is an abundance of speeches already recorded and published that needs
to be transcribed, and more data are being generated
every day~\cite{asr_largedata}. Hence, it is important to recognize large
speech databases very quickly at the lowest cost. In this work, we present
efficient methods to utilize the Transformer in continuously spoken long speech
recognition.

\subsection{Literature review}

In 2017, a deep learning model, Transformer, was introduced for machine
translation tasks~\cite{transformer}. Transformer has been adopted in many
natural language processing tasks and produced significant performance
improvements. In 2018, Transformer was adopted for speech recognition tasks with
significant improvements when compared to recurrent neural network
(RNN)-based speech recognition~\cite{speech_transformer, transformer_asr_201806,
transformer_asr_201809}. Accordingly, there have been many studies on
Transformer-based end-to-end speech recognition, which achieved state-of-art
speech recognizers~\cite{espnet_transformer2019, conformer, convtransformer_transducer}. Moreover,
\cite{joint_ctc_transformer} successfully
integrated connectionist temporal classification (CTC) into Transformer-based end-to-end speech recognition for faster training,
and subsequently improved the speech recognition performance. Recently, a joint
CTC\slash{}Transformer approach was used in various
research studies~\cite{espnet_transformer2019, Moritz20,
online_ctc_att_asr2020, multispk_asr_w_ctc2020, nonar_w_ctc2020,
insertion_based_modeling_w_ctc2020, e2e_sincnet2020, self_mixed_att_2020,
nonar_w_ctc2020v2, self_dist_2020, long_context_2020, cross_att_2020,
code_switching_2020, adapt_adjust_2020, lightweight_2020,
unsup_pretraining_mtkl_2020}. Therefore, this paper focuses on
CTC\slash{}Transformer-based end-to-end speech recognition.

To increase the throughput of RNN-based speech recognizers, 
\cite{deepspeech2, nvidiabatch, batchBLSTM} utilized several batch processing
approaches. In particular, \cite{nvidiabatch, batchBLSTM} aimed to
accelerate GPU parallelization. \cite{e2e_vector_beam_search2018,
e2e_vector_beam_search} proposed a multiple-utterance multiple-hypothesis
vectorized beam search in CTC-attention-based end-to-end speech recognition using a VGG-RNN-based
encoder-decoder and showed the decoding throughput increase using a GPU.
Motivated by~\cite{e2e_vector_beam_search2018, e2e_vector_beam_search}, we aim
to successfully adopt the vectorized beam search in
CTC\slash{}Transformer-based end-to-end speech recognition for faster and more efficient decoding
of a large speech database.

The CTC prefix decoding used in CTC\slash{}Transformer end-to-end speech recognition requires
computation over an entire input sequence for each iteration~\cite{ctc} and
is not parallel processing but sequential processing, which hinders the GPU
parallelization. Since this is a bottleneck in fast processing,
several time-restricted CTC prefix decoding
methods~\cite{e2e_vector_beam_search, restricted_ctc} have been proposed. 
\cite{e2e_vector_beam_search} proposed a windowing method using attention
weights of a VGG-RNN decoder network with the assumption that the time
alignment of the CTC prefix decoding is similar to that of the attention
decoding. \cite{restricted_ctc} proposed truncated CTC prefix
decoding using the previous and current CTC probabilities of a blank symbol at
each iteration. For fast decoding, we propose time-restricted CTC prefix
decoding via a different method.

Voice activity detection (VAD) is one solution for decoding long utterances
robustly with low resources in an end-to-end speech recognizer
\cite{Yoshimura20}. \cite{Yoshimura20} proposed a CTC-based VAD
integrated into an end-to-end speech recognizer and compared this to an energy-based VAD
method and time-delay neural network (TDNN)-based VAD. This
paper proposes a DNN-based VAD with a different
method for the vectorized beam search in CTC\slash{}Transformer end-to-end
speech recognition.

\subsection{Proposed Method and Its Contribution}

In this paper, we propose a method to recognize a large speech database very
quickly and cost-effectively for CTC\slash{}Transformer-based end-to-end speech recognition.
To this end, we adopt the multiple-utterance multiple-hypothesis
vectorized beam search decoding of \cite{e2e_vector_beam_search, restricted_ctc}
in Transformer-based speech recognition. We observed that the baseline end-of-speech
detection method fails if an abnormal utterance is encountered since the method
is based on the absolute value of CTC probability. Thus, we propose an
end-of-speech detection method using CTC-based time information.
To reduce the deterioration of GPU parallelization, CTC prefix decoding is moved
from the GPU to the CPU. To unburden the CPU processing load, we propose a
time-restricted CTC prefix decoding method using the previous CTC probabilities
of the target symbol and its blank symbol for each iteration because it is difficult to
determine the time information from a Transformer decoder, unlike
\cite{e2e_vector_beam_search, restricted_ctc}. Another issue is the
limited GPU memory size since batchfied utterances are padded with the maximum
utterance length before the vectorized beam search
\cite{e2e_vector_beam_search2018, e2e_vector_beam_search}. Therefore, we propose
a method to segment long speeches into smaller pieces for batch decoding,
and create a batch having speech segments of similar lengths.

The contributions of this paper are summarized as follows:
\begin{description}[style=unboxed]
 \item[Multiple-utterance multiple-hypothesis batched beam search for Transformer]
 As a prior work, multiple-utterance multiple-hypothesis batched decoding was
 utilized for a joint CTC/Attention-based VGG-RNN. Motivated by this 
 prior work, we extend the multiple-utterance batched decoding for
 Transformer-based speech recognition.

 \item[Time-restricted CTC prefix scoring for Transformer]
	Influenced by
 time-restricted CTC prefix scoring for a VGG-RNN-based model, we
 present time-restricted CTC prefix scoring using CTC-based time
 information instead of attention weights.

 \item[Improvement in end-of-speech detection]
	We propose an
 improved end-of-speech detection based on estimated time information.
 That is, conventional end-of-speech detection works well in matched
 conditions with speeches in training corpus,
 but if an abnormal speech is encountered, then the method could
 fail to detect end-of-speech. Thus, the decoding efficiency is
 degraded. To solve this problem, we improve conventional end-of-speech
 detection.

 \item[{\color{\BLUE}Utterance segmentation for efficient decoding with the} Transformer]
	In order to efficiently
 manipulate GPU-based batched beam search decoding for the Transformer, it
		is important to split utterances to be decoded. First, we {\color{\BLUE}adopted} a
 DNN-VAD based segmentation to split at pauses. Using this method,
 the decoding speed can be significantly accelerated with little accuracy degradation.
		Then, it is shown that hard segmentation can be {\color{\BLUE}also} used as a splitting
		method with minor {\color{\BLUE}accuracy} degradation. It is also analyzed that sorting and
 making batches of utterances of similar lengths helps to speed up the
 decoding. By integrating these techniques, we can deploy highly
 efficient end-to-end speech recognizers for real-world applications.
\end{description}

The remainder of this paper is organized as follows. {\color{\BLUE}Section~\ref{sec:review}
reviews a CTC\slash{}Transformer-based speech recognition in brief. In}
Section~\ref{sec:transformer2}, the methods we utilized to speed up the
recognition for large speech corpus are presented. Next, in
Section~\ref{sec:expr}, the performance of the proposed speech recognition is evaluated with
a public dataset and a real-world meeting corpus spoken by multiple speakers.
Finally, we conclude the paper with our findings in
Section~\ref{sec:conc}.

%% file: review.tex
{\color{\BLUE}
\section{CTC\slash{}Transformer-based speech recognition}
\label{sec:review}
The authors of \cite{transformer} propose a sequence-to-sequence
attention-based encoder-decoder network, Transformer, for a neural machine
translation task. With an outstanding performance, Transformer is
successfully adopted in speech recognition \cite{speech_transformer,
transformer_asr_201809, espnet_transformer2019} and this work also adopts the
Transformer network of \cite{espnet_transformer2019}. As shown in
Fig.~\ref{fig:transformer_asr}, the encoder network consists of sub-sampling,
input embedding, position encoding, $N_e$ encoder blocks, and normalization
layers. The
decoder network consists of output embedding, positional encoding, $N_d$ decoder
blocks, normalization, linear, and softmax layers. A linear layer and a softmax are
also used for a CTC decoder, which is shown in orange-line in the figure. Each
encoder block consists of multi-head attention and feed-forward layers. Each decoder
block consists of masked multi-head attention, multi-head attention, and 
feed-forward. Layer normalization is applied before each multi-head attention
and feed-forward and residual addition is applied after each of them. The
following subsections briefly explain a joint CTC\slash{}Transformer score and a
beam search decoding for speech recognition inference.

\begin{figure}[t]
    \centering
			\includegraphics[scale=0.5]{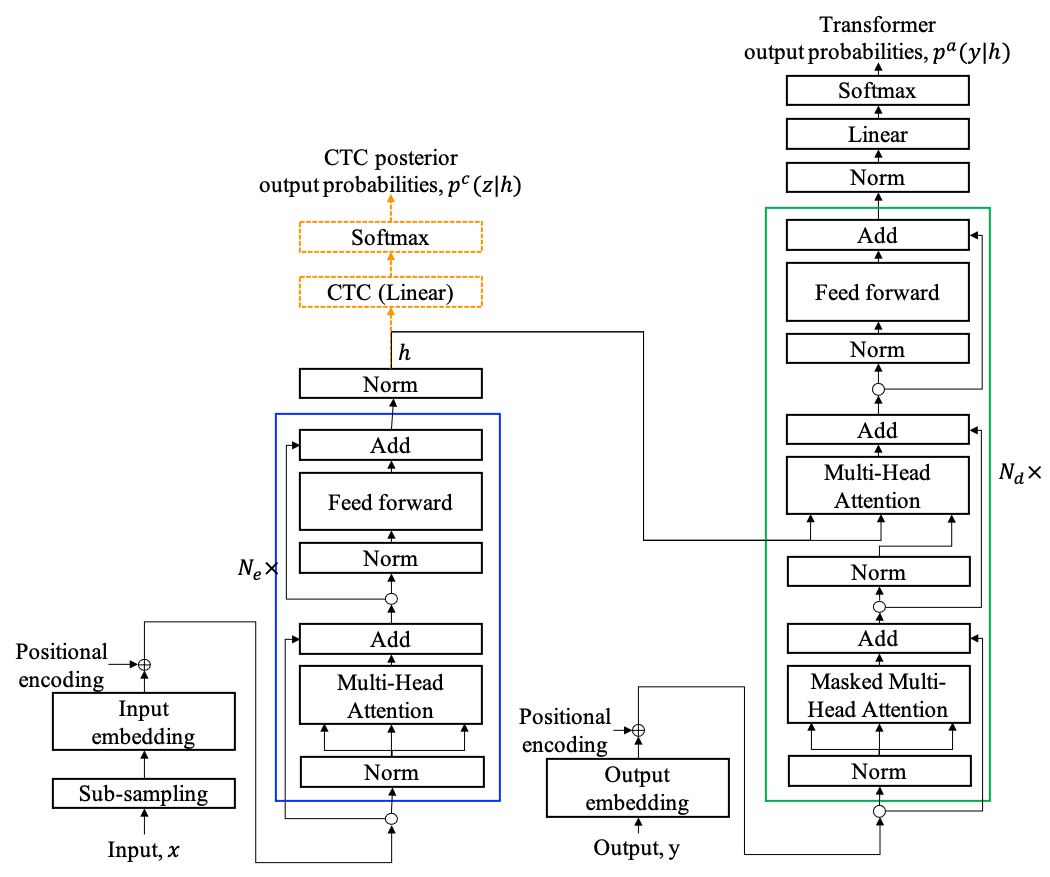}
	\caption{\color{\BLUE}Overall structure of the CTC\slash{}Transformer-based speech
	recognizer that is used in this work \cite{espnet_transformer2019}. Encoder
	blocks and decoder blocks are depicted in blue-line and in green-line,
	respectively. A CTC decoder is shown in orange-line.%
    \label{fig:transformer_asr}}
\end{figure}

\subsection{Joint CTC\slash{}Transformer score}

In~\cite{joint_ctc_transformer,e2e_vector_beam_search}, a joint
CTC\slash{}attention decoding is performed to maximize the logarithmic linear
combination of the sequence probabilities that are obtained by the Transformer
and CTC decoders. For the $l$-length sequence, $\textbf{y}^{1:l}$, of a speech
representation, $\textbf{h}$, this work calculates a joint
CTC\slash{}Transformer
score \cite{joint_ctc_transformer} as follows,
\begin{linenomath}
\begin{equation} \label{eq:joint_score}
    {P^{\textrm{joint}}} (\textbf{y}^{1:l}|\textbf{h}) = %
    \lambda {P^{\textrm{CTC}}} (\textbf{y}^{1:l}|\textbf{h}) + \left ( 1
    - \lambda \right ) {P^{\textrm{Att}}} (\textbf{y}^{1:l}|\textbf{h}),
\end{equation}
\end{linenomath}
where  ${P^{\textrm{CTC}}} (\textbf{y}^{1:l}|\textbf{h})$ and
${P^{\textrm{Att}}} (\textbf{y}^{1:l}|\textbf{h})$
are the CTC prefix probability and the
attention probability respectively. And $\lambda$ is the CTC weight.

The attention probability, ${P^{\textrm{Att}}}(\textbf{y}^{1:l}|\textbf{h})$, is
calculated as,
\begin{linenomath}
\begin{equation}
    {P^{\textrm{Att}}} (\textbf{y}^{1:l}|\textbf{h}) = %
    {P^{\textrm{Att}}} (\textbf{y}^{1:l-1}|\textbf{h}) + %
    p^{a}(\textbf{y}^{l}|\textbf{y}^{1:l-1},\textbf{h})
\end{equation}
\end{linenomath}
where $p^{a}(\textbf{y}^{l}|\textbf{y}^{1:l-1},\textbf{h})$ is the attention output probability at
time $l$ given $\textbf{y}^{1:l-1}$ and ${\textbf{h}}$, as shown in
Fig.~\ref{fig:transformer_asr}.

Next, CTC \cite{ctc} maps a frame-wise speech feature vector to a
token-wise output sequence in a monotonic manner. Let $\textbf{h}$  be a
frame-wise $T$-length speech representation and $\textbf{y}_{1:L}$ be a
token-wise $L$-length output sequence, $\textbf{y}_{1:L} = \left \{ y_l \in C |
l = 1, \cdots , L \right \}$, with a token set $C$.  The CTC introduces a
frame-wise $T$-length output sequence, $Z = \left \{ z_t \in \hat{C} | t = 0,
\cdots , T \right \}$, with an extended token set $\hat{C}= \left \{ C\cup
\mathrm{blank}  \right \}$. $z_t$ can be defined as,
\begin{eqnarray}
	\mathrm{blank,} && \textrm{if } t = 0  \\
	y^{l-1} \textrm{~or~} y^{l} \textrm{~or~}  \mathrm{blank,} && \textrm{if } z_{t-1} = y^{l-1} \textrm{ and } y^{l-1} \neq y^{l} \\
	y^{l-1} \textrm{~or~} \mathrm{blank,} && \textrm{if } z_{t-1} = y^{l-1} \textrm{ and } y^{l-1} =  y^{l} \\
	y^l \textrm{~or~}  \mathrm{blank,} && \textrm{if } z_{t-1} = \textrm{blank and } z^{nb}_{t-1} = y^{l-1}
\end{eqnarray}
where the first and last non-blank tokens are $y^1$ and $y^L$. ${z}^{nb}_{t}$ is
the previous non-blank token up to $t$. The best output sequence of
$Z$ is obtained by finding a sequence that maximizes a CTC prefix
probability \cite{joint_ctc_att,e2e_vector_beam_search,restricted_ctc}. The CTC
prefix probability accumulates the probabilities of all possible sequences
having a prefix sequence. That is, for an $l$-length hypothesis
$\textbf{y}^{1:l}$ of a speech representation ${\textbf{h}}$, the CTC prefix
probability, ${P^{\mathrm{CTC}}} (\textbf{y}^{1:l}|\textbf{h})$, is calculated as,
\begin{linenomath}
\begin{equation} \label{eq:ctc}
    \sum_{l \leq t \leq T}
		{\phi}_{{t}-1} (\textbf{y}^{1:l-1}|\textbf{h})p^{c}%
        (z_t=\textbf{y}^{l}|\textbf{y}^{1:l-1}, \textbf{h}),
\end{equation}
\end{linenomath}
where ${\phi}_{t-1}(\textbf{y}^{1:l-1})$ is the CTC forward probability up to
time $t-1$ for $\textbf{y}^{1:l-1}$. $p(z_t|\textbf{y}^{1:l-1},\textbf{h})$ is
the CTC posterior probability at time $t$ given $\textbf{y}^{1:l-1}$ and
${\textbf{h}}$, as shown in Fig.~\ref{fig:transformer_asr}.

\subsection{Beam search decoding}
This work uses a beam search decoding \cite{bestfirst_beamsearch} to find the
output sequence for an input utterance. The beam search is a beam-pruned
breadth-first search, as summarized in Algorithm~\ref{alg:beam_search}.  Let
$\textbf{x}$ be a speech feature. Using a Transformer encoder, $\textbf{x}$ is
converted into the intermediate representation $\textbf{h}$ where $\left |
\textbf{h} \right |$ is the length of $\textbf{h}$. Then, $\textbf{h}$ is
converted into the text sequence $\textbf{y}$ by performing the $B$-width beam
search where Eq.~\ref{eq:joint_score} is used as the scoring function to
evaluate the reliability of a hypothesis (line~3 of Algorithm~\ref{alg:beam_search}).

At the $l$-th step of the beam search, the $l$-length hypothesis scores,
${\mathbf{P}_{\left [ \left | C \right | \right ]}}$, are calculated using
$\textbf{h}$, $\mathbf{Y}^{l-1}$, and $C$ (line~7 of
Algorithm~\ref{alg:beam_search})). It is noted that the notation
${A}_{\left [ B \right ]}$ means that a matrix $A$ has the shape of $B$.
${\mathbf{I}}_{\left [ B \right ]}$ is obtained by selecting the indices of
the $B$ highest scores of ${\mathbf{P}_{\left [ \left | C \right | \right ]}}$
(line~10 of Algorithm~\ref{alg:beam_search}). Then, the $l$-length hypothesis
set, ${\mathbf{Y}}^{l}_{\left [ B \right ]}$, is obtained by selecting the
$l$-length reliable hypotheses using $\mathbf{Y}^{l-1}$, $C$, and
${\mathbf{I}}_{\left [ B \right ]}$ (line~11 of
Algorithm~\ref{alg:beam_search}). It is noted that $a \circ B$ indicates that $a$ is concatenated with each item of $B$.
Next, $\hat{\mathbf{Y}}$ is updated if
\emph{\textless eos\textgreater}-ended hypothesis is confident (lines~12--15 of
Algorithm~\ref{alg:beam_search}).

The beam search is terminated if the decoding step $l$ is greater than $ \left |
{\textbf{h}} \right |$ (line~4 of Algorithm~\ref{alg:beam_search}) or if
end-of-speech is detected (lines~17--19 of Algorithm~\ref{alg:beam_search}).
Finally, the output sequence is obtained with the most probable hypothesis in
$\hat{\textbf{Y}}$ (line~21 of Algorithm~\ref{alg:beam_search}).

\begin{algorithm}[t]
	\caption{\color{\BLUE}$B$-width beam search decoding}\label{alg:beam_search}
\begin{algorithmic}[1]
	\color{\BLUE}
	\Require {input feature representation $\textbf{h}$, beam width $B$}
	\Ensure {output sequence $y$}

		\State ${\mathbf{Y}}^0 \leftarrow \left \{ \emph{\textless eos\textgreater} \right \} $
		\State $\hat{\mathbf{Y}} \leftarrow \varnothing $
        \State score\_fn($\cdot$) $ \leftarrow {P^{\mathrm{joint}}} ( \cdot  | \textbf{h})  $ 

		\For{$l \in \left \{ 1, \cdots, \left | \textbf{h} \right | \right \}$}
			\State $ {\mathbf{Y}}^l \leftarrow \varnothing $
			\For{$ y   \in {\mathbf{Y}}^{l-1}$ }
				\State ${\mathbf{P}_{\left [ \left | C \right | \right ]}} \leftarrow$ score\_fn( $y$ )
				\State ${ \hat{\textrm{P}}} \leftarrow {\mathbf{P}_{\left [ \left | C \right | \right ]}} \left ( \emph{\textless eos\textgreater} \right )$
				\State $ {\mathbf{P}}_{\left [ \left | C \right | \right ]} \left ( \emph{\textless eos\textgreater} \right ) \leftarrow$ log0

				\item[]
				\State $ {\mathbf{I}}_{\left [ B \right ]}  \leftarrow$ topB( $ {\mathbf{P}}_{\left [ \left | C \right | \right ]} $ )
				\State $ {\mathbf{Y}}^{l}_{\left [ B \right ]} \leftarrow$ gatherB( $y \circ C$, $\mathbf{I}_{\left [ B \right ]}$ )

				\item[]
				\State best\_score = ${max}{\mathbf{P}_{\left [ \left | C \right | \right ]}}$
				\If { ${ \hat{\textrm{P}}}$ > best\_score }
                    \State $\hat{\mathbf{Y}}.\mathrm{add}$( $ \left ( y \circ \emph{\textless eos\textgreater}, { \hat{\textrm{P}}} \right ) $ )
				\EndIf

			\EndFor

			\If{ ENDDETECT($\hat{\mathbf{Y}}$, $l$) == true }
					\State $\textbf{break}$
			\EndIf
		\EndFor

			\State \Return $ \argmax_{(y,p) \in \hat{\mathbf{Y}} } p  $

\end{algorithmic}
\end{algorithm}
}

%% file: batch.tex
\subsection{Fast decoding based on batched beam search}
\label{sect:batch}

GPU parallelization can be accelerated by batch processing~
\cite{e2e_vector_beam_search, batchBLSTM, e2e_vector_beam_search2018,
deepspeech2,nvidiabatch}. As in \cite{e2e_vector_beam_search,
e2e_vector_beam_search2018}, we adopt a multiple-utterance multiple-hypothesis batched beam search for a
Transformer-based end-to-end speech recognizer to efficiently parallelize speech
recognition for multiple utterances. The batched beam search is briefly
described in the following, and the detailed explanation is in
\cite{e2e_vector_beam_search}.

{\color{\BLUE}{
Let $\textbf{X}=\left \{\textbf{x}_{1}, \cdots, \textbf{x}_{U} \right \}$ be the speech
features that are obtained from $U$ utterances. $\textbf{X}$ is
batched as $\mathbb{{X}} = \left \{ {\textbf{x}}_{1},
\cdots, {\textbf{x}}_{U} \right \}$ where ${\textbf{x}}_{i}$ is length-extended from 
${x}_{i}$. The length is $ {\max}_{\textbf{x}_i \in \textbf{X}}{\left | \textbf{x}_i \right |}$
and the extended values are padded with zeroes.  Using a Transformer encoder,
$\mathbb{{X}}$ is converted into the intermediate representations $\mathbb{H} =
\left \{ {\textbf{h}}_{1}, \cdots, {\textbf{h}}_{U} \right \}$, where each of
them can be converted in parallel. Then, $\mathbb{H}$ is decoded into the text
sequence set $\mathbb{Y} = \left \{ {\textbf{y}}_{1}, \cdots, {\textbf{y}}_{U}
\right \}$ by performing a $B$-width batched beam search. For the scoring
function, Eq.~\ref{eq:joint_score} is used to evaluate each hypothesis of
$\mathbb{{H}}$ in parallel (line~3 of Algorithm~\ref{alg:batched_beam_search}).

At the $l$-th step of the batched beam search, the score set,
${\mathbb{P}}_{\left [ U \times B, \left | C \right | \right ]}$, of
$B$ $l$-length hypotheses of $U$ speech features is calculated using
$\mathbb{H}$, $\mathbb{Y}^{l-1}$, and $C$, where each score can be calculated in
parallel (line~6 of Algorithm~\ref{alg:batched_beam_search}).
${\mathbb{I}}_{\left [ U ,  B \right ]}$ is obtained by selecting the
indices of the $B$ highest scores of the $U$ speech feature using
${\mathbb{P}}_{\left [ U \times B, \left | C \right | \right ]}$ (line~9 of
Algorithm~\ref{alg:batched_beam_search}). Then, the $l$-length hypothesis set,
${\mathbb{Y}}^{l}_{\left [ U \times B \right ]}$, is obtained by selecting the
$B$ $l$-length reliable hypotheses for each speech feature using
$\mathbb{Y}^{l-1}$, $C$, and ${\mathbb{I}}_{\left [ U \times B,  B \right ]}$
(line~10 of Algorithm~\ref{alg:batched_beam_search}). Next, confident
\emph{\textless eos\textgreater}-ended hypotheses for each utterance are added
into $\hat{\mathbb{Y}}\left ( i \right )$ (lines~11--18 of
Algorithm~\ref{alg:batched_beam_search}).

The beam search is terminated if the decoding step $l$ is greater than $\mathrm{len}_{h}
= {\max}_{ h_i \in \mathbb{H} }{ \left | h_i \right | }$ (line~4 of
Algorithm~\ref{alg:batched_beam_search}) or if end-of-speech is detected for all
utterances (lines~19--22 of Algorithm~\ref{alg:batched_beam_search}). Finally,
the output sequence is obtained with the most reliable hypothesis in
$\hat{\mathbb{Y}}$ for each utterance (line 24 of
Algorithm~\ref{alg:batched_beam_search}).

\begin{algorithm}[t]
	\caption{\color{\BLUE}$B$-width batched beam search decoding}\label{alg:batched_beam_search}
\begin{algorithmic}[1]
	\color{\BLUE}
	\Require {input feature representation set $\mathbb{H} = \left \{ {\textbf{h}}_{1}, \cdots, {\textbf{h}}_{U} \right \}$, beam width $B$}
	\Ensure {output sequence set $\mathbb{Y} = \left \{ {\textbf{y}}_{1}, \cdots, {\textbf{y}}_{U} \right \}$}

		\State ${\mathbb{Y}}^0_{\left [ U \times B \right ]} \leftarrow 
			\left \{ \left \{ \emph{\textless eos\textgreater} \right \}, \cdots,  \left \{ \emph{\textless eos\textgreater} \right \} \right \}$
		\State $\hat{\mathbb{Y}}_{\left [ U \right ]}  \leftarrow 
			\left \{ \left \{ \varnothing \right \}, \cdots, \{ \left \{ \varnothing \right \}  \right \}$
		\State Score\_Fn($\cdot$, ${\mathbb{H}}$) $\leftarrow 
			{P^{joint}} ( \cdot  | \mathbb{H} ) $

		\For{$l \in \left \{ 1, \cdots, 
			{max}_{ h_i \in \mathbb{H} }{ \left | h_i \right | }	
			\right \}$}	
			\State ${\mathbb{Y}}^{l}_{\left [ U \times B \right ]}  \leftarrow 
				\left \{ \left \{ \varnothing \right \}, \cdots, \{ \left \{ \varnothing \right \}  \right \}$

			\item[]
			\State $ {\mathbb{P}}_{\left [ U \times B, \left | C \right | \right ]} \leftarrow$
				Score\_Fn( $ \mathbb{Y}^{l-1}$ ) 
			\State ${ \hat{\mathbb{P}}}_{\left [ U \times B \right ]} \leftarrow {\mathbb{P}} \left ( \cdot, \emph{\textless eos\textgreater} \right ) $
			\State $ {\mathbb{P}}_{\left [ U \times B, \left | C \right | \right ]} \left ( \cdot, \emph{\textless eos\textgreater} \right ) \leftarrow$ log0

			\item[]
			\State $ {\mathbb{I}}_{\left [ U ,  B \right ]}  \leftarrow$ TopB( $ {\mathbb{P}}_{\left [ U \times B, \left | C \right | \right ]} $ )
			\State $ {\mathbb{Y}}^{l}_{\left [ U \times B \right ]} \leftarrow$ GatherB( $\mathbb{Y}^{l-1}_{\left [ U \times B \right ]}$, $C$,
			$\mathbb{I}_{\left [ U,  B \right ]}$ )  

			\item[]
				\State $\textbf{best\_score}$ = $\left \{
					{max}_{ j \in \left | C \right | } {\mathbb{P}}\left ({i, j} \right )  | 1 \leq i \leq U
					\right \}$
			\For{$ i \in \left \{ 1 \cdots U \right  \} $ and $ j \in \left \{ 1 \cdots B \right \} $}
			\State $p = \hat{\mathbb{P}}{\left ( i \times U + j \right )} $
			\State $y = \mathbb{Y}^{l-1} \left ( i \times U + j  \right )  \circ \emph{\textless eos\textgreater}$
					\If { $p$ > $\textbf{best\_score}$($i$) }
                        \State $\hat{\mathbb{Y}}\left ( i \right )%
                        .\mathrm{add}$( $ \left ( y, p \right ) $ )
					\EndIf
			\EndFor

			\item[]
				\State is\_eos = $\left \{ \textrm{ENDDETECT(}\hat{\mathbb{Y}}\left ( i \right ), l) | 1 \leq i \leq U \right \}$

			\If{ not ( false $\in$ is\_eos ) }
					\State $\textbf{break}$
			\EndIf
		\EndFor

		\item[]
			\State \Return $ \left \{ 
                \argmax_{(y,p) \in \hat{\mathbb{Y}}(i) } p | 1 \leq i \leq U
			\right \}$

\end{algorithmic}
\end{algorithm}

When compared to \cite{e2e_vector_beam_search, e2e_vector_beam_search2018}, this
work uses the different scoring function for a CTC\slash{}Transformer-based
speech recognition (line~3 of Algorithm~\ref{alg:batched_beam_search}) and
improves an end-of-speech detection based on CTC (line~19 of
Algorithm~\ref{alg:batched_beam_search}). For a fast decoding, this work also
proposes a time-restricted CTC scoring of the scoring function.
}}

%% file: ctc.tex
\subsection{CTC-based end-of-speech detection and time-restricted CTC prefix scoring}
\label{sect:ctc}

This section first explains {\color{\BLUE}{a baseline end-of-speech detection
\cite{hybrid_ctc_att} and then proposes fast decoding methods based on a
CTC prefix score.}} 

\subsubsection{{\color{\BLUE}{Baseline end-of-speech detection}}}
As mentioned in Section  \ref{sect:batch}, the beam search decoding finishes
when a confident \emph{\textless eos\textgreater}{\color{\BLUE}{-ended hypothesis}} is detected for a fast
completion. Based on \cite{hybrid_ctc_att},
the baseline end-of-speech detection detects {\color{\BLUE}{end-of-speech if 
the following equation is satisfied,
\begin{linenomath}
\begin{equation} \label{eq:ctc_eos_base}
	\sum_{m=0}^{M-1}{\left [ \left \{ 
	\underset{y \in \hat{\mathbb{Y}}(i):|y|=l-m}{max}{P(y|{\textbf{h}}_{i}}) - 
	\underset{y^{'} \in \hat{\mathbb{Y}}(i)}{max} {P(y^{'}|{\textbf{h}}_{i}})
	\right \} < D_{end} \right ]} = M.
\end{equation}
\end{linenomath}}}
In this work, the predefined parameters $M$ and
$D_{end}$ are set to 3 and $e^{-10}$, as in \cite{hybrid_ctc_att}.  

\subsubsection{Proposed CTC-based end-of-speech detection}

The baseline end-of-speech detection of \cite{hybrid_ctc_att} performs well in a
normal condition. However, the absolute probability values of hypotheses are
unreliable in a mismatched condition
between training and decoding, which can occur in real-world
applications. To prevent such a malfunction that slows down speech recognition, we
propose a CTC-based end-of-speech detection method that uses the estimated time of
the last label (\emph{\textless eos\textgreater}) of each completed hypothesis.

\begin{figure}[t!]
	\caption{Example of time information for each decoded output unit using estimated start time index.\\}
    \label{fig:time-info}
    \centering
		\subfloat[{\color{\BLUE}{Illustration of the start time indices}}]{
			\includegraphics[width=0.9\columnwidth]{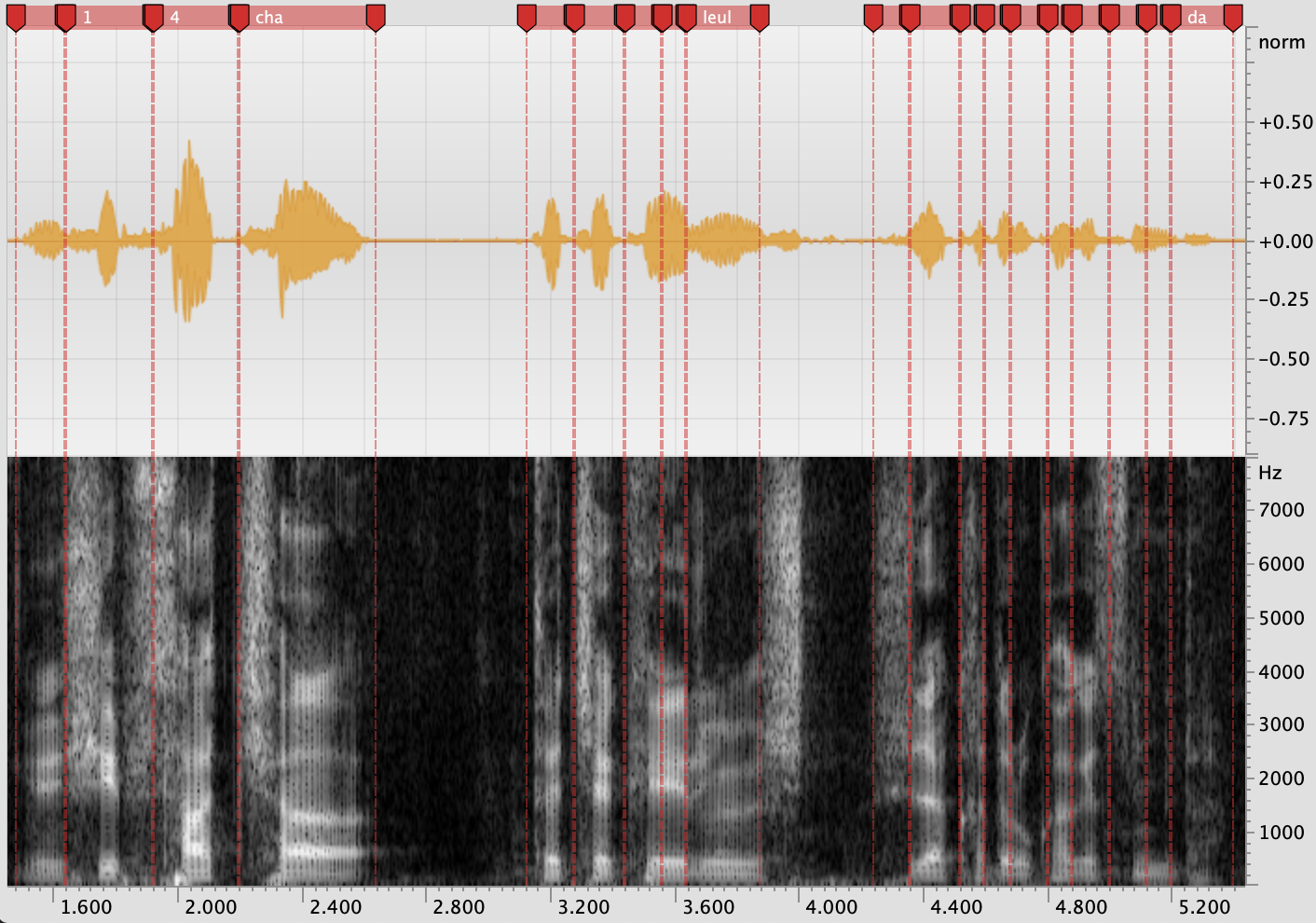}
			\label{fig:time-info:1}
				}\\
		\subfloat[{\color{\BLUE}{Start time indices for a wrong hypothesis having repeatition errors. $len_h$ is 100 in this example.}}]{
			\includegraphics[width=0.9\columnwidth]{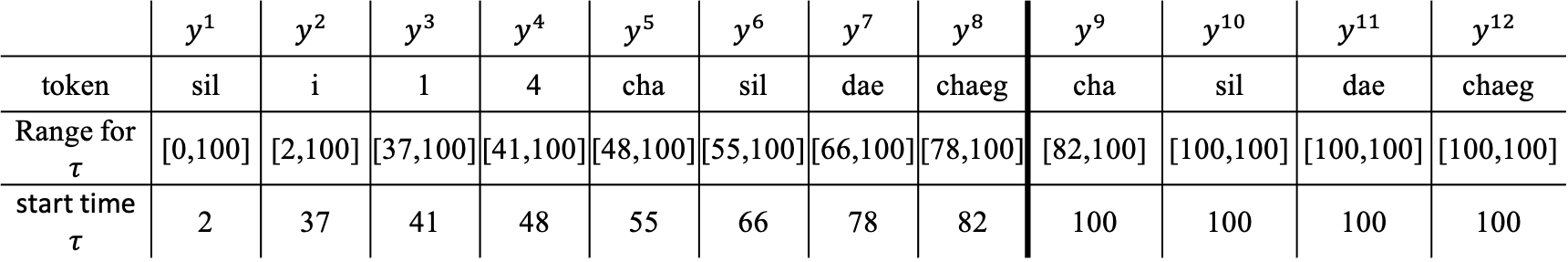}
			\label{fig:time-info:2}
				}
\end{figure}

Based on a CTC prefix score, {\color{\BLUE}{for the $j$-th $l$-length hypothesis, $\mathbf{y}_{i,j}^{1:l}$, of an $i$-th feature representation, }}the start time index
$\tau_{i,j}^{l}$ and end time index ${\tilde\tau}_{i,j}^{l}$ {\color{\BLUE}{of}} the last
label{\color{\BLUE}{, $\mathbf{y}_{i,j}^{l:l}$,}} are estimated as follows:
\begin{linenomath}
\begin{eqnarray}
	{\tau}_{i,j}^{l} &=& \argmax_{\tau_{i,j}^{l-1} \leq t \leq {\color{\BLUE}{{len_{h}}}} }
		{\phi}_{{t}}(\mathbf{y}_{i,j}^{{\color{\BLUE}{1:l}}}) \\
		{\tilde\tau}_{i,j}^{l} &=& \argmax_{\tau_{i,j}^{l-1} \leq t \leq {\color{\BLUE}{len_{h}}}}
		{\phi}_{{t}}(\mathbf{\tilde{y}}_{i,j}^{{\color{\BLUE}{1:l}}}) 
\end{eqnarray}
\end{linenomath}
where $\mathbf{\tilde{y}}$ is a text sequence that ends with a blank symbol
concatenated after ${\mathbf{y}}$. That is, $\tau_{i,j}^{l}$ is a time index
that maximizes the CTC forward probability for 
{\color{\BLUE}{$\mathbf{y}_{i}^{1:l}$}} within a specified time index range. Similarly,
${\tilde\tau}_{i,j}^{l}$ is one for $\mathbf{\tilde{y}}$.
As shown in {\color{\BLUE}Fig.~\ref{fig:time-info:1}}, we validate the estimated time index by
obtaining the start time of the decoded output text for a Korean utterance using the
proposed start time index. Note that the upper and lower parts show the
waveform in yellow and its spectrum in gray for the utterance.
The estimated start time is shown in red. It can be seen that the start
time is nearly estimated at the start position of each syllable.
{\color{\BLUE}On the other hand, Fig.~\ref{fig:time-info:2} shows an example of the start time
indices for a wrong hypothesis having repeatition errors, which errors can be
occurred in an attention-based speech recognition. As shown in the figure, the
repeatition errors are $\left \{ y^9, \cdots, y^{12} \right \}$ and the start
time indices for the errors are same as $len_h$=100 due to the limited search
range of $\tau$.}

{\color{\BLUE}{Based on the property that is depicted in Fig.~\ref{fig:time-info:2}, the
CTC-based end-of-speech detection method is proposed, as in
Algorithm~\ref{alg:ctc_eos}. If the baseline end-of-speech detection method
fails to predict end-of-speech, the start time index is examined for each
hypothesis in the confident \emph{\textless eos\textgreater}-ended hypothesis
set $\hat{\mathbb{Y}}(i)$ for ${\textbf{h}}_{i}$. If the start time index is
equal to $len_h$ more than $C$ times, end-of-speech is detected.  In this study,
the parameter $C$ was set to 2.

\begin{algorithm}[t]
	\caption{\color{\BLUE}Proposed CTC-based end-of-speech detection}\label{alg:ctc_eos}
\begin{algorithmic}[1]
	\color{\BLUE}
	\Require {output sequence candidate set $\hat{\mathbb{Y}}(i)$, maximum length $len_{h}$}
	\Ensure {whether end-of-speech is detected}
	\If {Eq. \ref{eq:ctc_eos_base} is satisfied }
		\State \Return true
	\EndIf

	\State {count\_long\_hyp = 0}
	\For { $ y_{i,j}^{1:l} \in \hat{\mathbb{Y}}(i) $ }
		\If { ${\tau}_{i,j}^{l} = len_{h}$   }
			\State { count\_long\_hyp ++ }
		\EndIf
		\If {count\_long\_hyp > C }
			\State \Return true
		\EndIf
	\EndFor

	\State \Return false
\end{algorithmic}
\end{algorithm}
}}

\subsubsection{Proposed time-restricted CTC prefix score for fast decoding}

Since the sequential calculation of a CTC prefix score limits parallel
processing, we propose a CPU-based and time-restricted CTC prefix score for a joint
CTC\slash{}Attention based Transformer to improve computational efficiency.
{\color{\BLUE}{The calculation of the CTC prefix score of Eq.~}}
(\ref{eq:ctc}) is restricted within the proposed time range, as in \cite{e2e_vector_beam_search}. The proposed
time-restricted CTC prefix score is defined as follows:
\begin{linenomath}
\begin{equation} \label{eq:restrictedctc}
    \sum_{s_{i,j}^{l} \leq t \leq e_{i,j}^{l}}%
	{\phi}_{{t}-1}(\mathbf{y}_{i,j}^{\color{\BLUE}1:l-1}{\color{\BLUE}{|{\textbf{h}}_{i}}})p^{\color{\BLUE}c}(z_t=\mathbf{y}_{i,j}^{\color{\BLUE}l:l}|{\color{\BLUE}{\mathbf{y}_{i,j}^{1:l-1},}}{\color{\BLUE}\textbf{h}}_{i}),
\end{equation}
\end{linenomath}
where $s_{i,j}^{l}$ and $e_{i,j}^{l}$ indicate the start and end times to be
calculated, respectively. These are calculated as
\begin{linenomath}
\begin{eqnarray}
	s_{i,j}^{l}= \max( \tau_{i,j}^{l-1} - \mathrm{M_1},\, l,\, 1), \\
	e_{i,j}^{l}= \min({\tilde\tau}_{i,j}^{l-1} + \mathrm{M_2},\, \left | {\color{\BLUE}h}_{i} \right |),
\end{eqnarray}
\end{linenomath}
where $M_1$ and $M_2$ are tunable margin parameters. For the batch processing of
${\mathbb{Y}}^{l}$, the restricted range is defined as
\begin{linenomath}
\begin{eqnarray}
    s^{l} &=& \min_{1 \leq i \leq U,\, 1 \leq j \leq B}s_{i,j}^{l}, \label{eq:starttime}\\
		e^{l} &=& \max_{1 \leq i \leq U,\, 1 \leq j \leq B} e_{i,j}^{l}, \label{eq:endtime}
\end{eqnarray}
\end{linenomath}
{\color{\BLUE}{The time-restricted CTC prefix score can be computed in a small amount of the
calculation. For a small but frequent calculation in a GPU device, memory
transfer load between CPU and GPU can be a bottleneck \cite{batchBLSTM}.
Therefore, the calculation of the CTC prefix score is moved from the GPU device to a
CPU device to reduce the GPU memory transfer load.}}

The proposed time-restricted CTC prefix score reduces the amount of computation
based on the
proposed estimated time information since a Transformer decoder
hardly captures time information. In contrast, \cite{e2e_vector_beam_search}
restricted the CTC prefix score using time information from
a RNN-based attention-decoder. \cite{restricted_ctc} only limited the
end time based on a differently estimated end time.

%% file: segment.tex
\subsection{Offline recognition of long utterances\label{sec:longutts}}

When it comes to the recognition of long utterances, the performance of
Transformer-based end-to-end speech recognition tends to degrade significantly owing to the
characteristics of the self-attention of the Transformer and the sensitivity to
the utterance length of training data \cite{relativePE_transformer}. On the
other hand, the computational complexity for a self-attention quadratically increases
in proportion to the utterance length~\cite{transformer,reformer}.
To tackle these issues, we perform segmentation before recognition. Two
segmentation methods are tested. The first is a split at a short pause with a
VAD, and the second is simple hard segmentation.

In \cite{Yoshimura20}, VAD information is used as a triggering sign for the
decoder in real time. In this work, explicit segmentation is performed
beforehand using a {\color{\BLUE}DNN-based VAD.

In conventional DNN-HMM ASR systems coupled with the language models, the
recognition results are searched by back-tracking the optimal path using the
Viterbi algorithm.
Usually the back-tracking point is determined as an end of utterance.
However, for long speeches, finding the relevant short pauses and using them as
back-tracking points makes it possible to generate ASR results quickly and hence
to reduce the user perceived latency.

Finding short pauses or detecting voice activity in continuous speech is
a difficult task and one of the oldest topics in speech-related researches.
Traditionally the methods to track the energies of speech and non-speech
are commonly used. These methods suffer when signal-to-noise ratio(SNR) is
dynamically changes.
In authors' previous work, outputs of
acoustic model (AM) have been successfully utilized to detect
voice activity in DNN-HMM based ASR systems.

In DNN-HMM based ASR systems,
a deep neural network is trained so that
each output node estimates a posteriori probability of a state given}
an input feature vector $\mathbf{x_t}$ at time $t$, $P(s_k|\mathbf{x_t})$,
where a state $s_k$ corresponds to the part of a phoneme.
{\color{\BLUE}The probabilities are directly used as the AM scores which
are combined with language model scores to find optimal ASR results.

In previous work, these probabilities are successfully used
to compute probabilities of voice presence in input signal as following.}

Let $o_i^t$ be the output value of the $i$-th node of the
neural network at time $t$. Then, speech presence probability ($P_\mathcal{S}(t)$)
and speech absence probability ($P_\mathcal{N}$) at time $t$ are estimated as
\begin{eqnarray}
    \log P_\mathcal{S}(t) &\approx& \max_{k\in \mathcal{S}} o_k^t \\
    \log P_\mathcal{N}(t) &\approx& \max_{k\in \mathcal{N}} o_k^t \\
    \mathrm{LLR}(t) &=& \log P_\mathcal{N}(t) - \log P_\mathcal{S}(t)
\end{eqnarray}
where $k\in\mathcal{S}$, and $k\in\mathcal{N}$ are output nodes corresponding to
the speech and noise states, respectively. A frame is regarded as a non-speech
frame if the log-likelihood ratio $\mathrm{LLR}(t)$ is larger than a given
threshold.  To obtain a more robust decision, the frame-wise results are smoothed
over multiple frames, as in \cite{batchBLSTM}.

{\color{\BLUE}We re-used DNN-based VAD algorithm to segment long speeches for
Transformer-based end-to-end speech recognition system.}
Segmentation based on DNN-VAD produces good performance in splitting long
utterances in general; however, this requires significant computation resources.
{\color{\BLUE}In DNN-HMM based ASR systems, these probabilities are computed for
ASR and there are no extra computation for VAD.
But for the Transformer-based end-to-end speech recognition system, there is
no such component like AM and this computation is only for the VAD, which
requires heavy computation with GPU for large DNN models.}

It is also possible to use the simpler VAD algorithm as in \cite{epd}.
In this work, we attempted hard segmentation by segment length, i.e., splitting
long utterances into short pieces of the same length while not considering
truncation. Hard segmentation requires no computation, and the resulting
segments are of the same length. This reinforces the merits of the batch
processing in Section \ref{sect:batch}.

%% file: baseline.tex
\subsection{Baseline ASR system and speech corpus}
\label{sec:baseline}

\subsubsection{Transformer-based end-to-end model}
\label{sec:transformer}
The Transformer model of the end-to-end speech recognizer used in this study
was trained using ESPnet, an end-to-end speech processing toolkit~\cite{espnet},
and the decoding was performed using the modified version of the toolkit. During the decoding,
an attention cache was applied in all experiments for speedup \cite{transformer_cache}.

As an input feature, 80-dimensional log-Mel filterbank coefficients were extracted for every 10--ms
analysis frame. The pitch feature was not used because in the Korean
language, tone and pitch are not important features. During training and
decoding, a global cepstral mean and variance normalization was applied to the
feature vectors.

Most hyperparameters for the Transformer model, inherit values from the
default settings of
the toolkit. The encoder consists of two convolutional layers, a linear
projection layer, and a positional encoding layer followed by 12 self-attention
blocks with layer-normalization. An additional linear layer was utilized for
CTC. The decoder has six
self-attention and encoder-decoder attention blocks. For every Transformer
layer, we used 2048-dimensional feedforward networks and four attention heads
with 256 dimensions.  The training was performed using the Noam
optimizer~\cite{transformer}, warmup steps, label smoothing,
gradient clipping, and accumulating gradients~\cite{accum_grad}.
100 epochs of training were performed without early stopping.

The text tokenization was performed in terms of character units. In Korean, a
character consists of two or three graphemes and corresponds to a syllable.  Although the
number of plausible characters in Korean is greater than 10,000, only 2,273 tokens
are used as output symbols, including digits, alphabets, and a spacing symbol, which
are seen in a training corpus. No other text processing was performed except for
the removal of punctuation marks and tokenization.

\subsubsection{Speech Corpus}
\label{sec:corpus}
To train the end-to-end ASR model, we utilized approximately 12.4k hours of Korean
speech corpus, which consist of a variety of sources including read digits and
sentences, recordings of spontaneous conversation, and mostly (about 11.1k
hours) broadcast data~\cite{Bang2020}. Therefore, the training data contain various utterance
lengths. All sentences were manually transcribed, and long sentences were
automatically segmented using the transcription and aligning process.
Utterances longer than 30~s were excluded when training.

As for the test corpus, meetings at a public office were recorded. An
automatic meeting transcription system was introduced for evaluation
purposes. In total, 8~h of recording were made from seven meetings. In each meeting,
4--22 people participated, and every participant had his or her own gooseneck-type
microphone. Each microphone was turned on only while the corresponding speaker
uttered,
and this was recorded separately. However, there were some overlaps and crosstalk from
adjacent speakers, which were ignored in the manual transcription. Each recording
lasted from 5~s to 36~min and was manually transcribed and split
according to the content by human transcribers. Each segment was of
0.6--42.9~s. The number and average lengths of segments are
shown in the first row of Table~\ref{tab:segmentinfo}.

%% file: exp_intro.tex
We performed ASR experiments to verify our proposed methods using two corpora:
the Librispeech corpus, a well-known English corpus; and
a real-world test dataset and the large training corpus explained in
Section~\ref{sec:corpus}. 

For experiments with the Librispeech corpus, one of the best models
published by ESPnet group was downloaded and used for decoding~\footnote{%
A Transformer model was retrieved from
\texttt{espnet/egs/librispeech/asr1/RESULTS.md}}. The speech recognition
performance in word error rate (WER) and the recognition speed are measured as
a real-time factor (xRT), which is the elapsed time divided by the utterance length,
to recognize a total 10.7~h of test speeches.

For the Korean test set, we trained a Transformer as described in
Section~\ref{sec:baseline}. The accuracy was measured as character error rates
(CER) instead of the word error rate (WER) because of ambiguity in the word
spacing. The recognition speed was measured as an xRT to
recognize 8~h of the test set. For both sets, the xRTs were averaged
after three trials.

All experiments in this section were performed on a workstation with two Xeon
{\color{\BLUE}Gold 5122 }CPUs and two GPU cards of a GeForce RTX 2080 Ti, which has 11.0 GB of GPU memory.

%% file: exp_batch.tex
\subsection{Experiments for batch decoding}
\label{subsec:batch-decoding}
First, experiments to verify the effect of batch decoding are performed.
As a baseline, we used the Transformer-based end-to-end ASR of
Section \ref{sec:transformer} and performed a single-utterance multiple-hypothesis $B$-width beam
search decoding on two GPUs. At each
decoding step, the $B(=3)$ hypotheses were batched, and the probabilities were
computed in parallel.
Multiple-utterance multiple-hypothesis batched decoding considers different lengths of
multiple utterances. However, the baseline ASR does not because it performs one utterance at a time.

Experimental results with Librispeech are listed in Table~\ref{tab:libri}.
Batchsize is set to 16 to fit in the GPU memory. As shown in the table,
multiple-utterance multiple-hypothesis batch decoding could recognize utterances
5.9 times faster with the same accuracy. {\color{\BLUE}As for two cases of the baseline and
the baseline without language model(LM) with narrow beam, the xRTs are not presented
because it was impractical to measure them. It took more than a few hours
to decode 10.7 hours of test data,
since the baseline system is not optimized for the fast decoding at all.

The accuracy of baseline with no LM and narrow beam should be same as that of
ours when batchsize is one, but it is slightly different; for test-clean 3.94
vs 3.92, and for test-other 8.63 vs 8.77. This is because the feature extraction
module is slightly different partly because of implementation issues and also
because of the dithering process in feature extraction. Dithering is used to
prevent zeros in the filterbank energy computation by adding random noises to
input signal.

The accuracy fluctuation depending on the batchsize (third and forth rows in
Table~\ref{tab:libri} is induced from the zero padding whiling making a batch.
To make a batch of multiple utterances, the zeros are padded to shorter inputs
to make the same length for the all utterances in the same batch.  Padded zeros
are masked by end-of-sentence(eos) detection algorithm, but still the zeros at
eos boundaries affects the computation and hence results in slightly different
ASR results.
}

\begin{table}[t!]
	\caption{Performance of baseline Transformer-based ASR employing 3-width single-utterance multiple-hypothesis beam search, and proposed
	ASR employing 3-width multiple-utterance multiple-hypothesis beam search on Librispeech
    evaluation set and model trained on 960-h trainset.  Experiments are
    performed using two GPUs. Total utterance duration is 10.7~h.}
\label{tab:libri}
\centering
\begin{tabular}{l|ccc}
    Configuration & \makecell{test\_\\[-0.5ex]clean} &
    \makecell{test\_\\[-0.5ex]other} & \makecell{xRT} \\  \hline
		Baseline & {\color{\BLUE}3.29} & {\color{\BLUE}{6.89}} & - \\
		No LM and narrow beam(=3.0) &  {\color{\BLUE}{3.94}} & {\color{\BLUE}8.63} & -\\ \hline
    Ours, batchsize=1 & 3.92 & 8.77 & 4.75e-2 \\ %
    Ours, batchsize=16 & 3.93 & 8.75 & 7.99e-3 \\ %
\end{tabular}
\end{table}

As for the Korean real-world test set,
the baseline ASR achieves a CER
of 9.06\% and average xRT of 0.256, as shown in Table~\ref{tab:batch_decoding}.
As a comparison, our previously developed conventional DNN-HMM based ASR system
using five-layer bidirectional long short-term model (bi-LSTM)
trained with the same training corpus
achieved a CER of 14.72\% with an external LM.

\begin{table}[t!]
	\caption{Performance of baseline ASR and proposed algorithm employing
    multiple-utterance batched beam search,
	CTC-based end-of-speech detection, and restricted CTC prefix score.
    Real-world Korean speech data are used.}
\label{tab:batch_decoding}
\centering
\begin{tabular}{l|cc}
	Method                               & CER(\%) & \makecell{xRT} \\ \hline
baseline                                 & 9.06    & 2.56e-1        \\ \hline %
+ batched decoding                       & 9.03    & 1.88e-2        \\        %
+ CPU CTC decoding                       & 9.06    & 1.15e-2        \\        %
+ CTC end-of-speech detection            & 9.08    & 1.05e-2        \\ \hline %
+ restricted CTC, $M_1$=5,$M_2$=$\infty$ & 9.07    & 9.11e-3        \\        %
+ restricted CTC, $M_1$=5,$M_2$=$60$     & 9.08    & 7.31e-3        \\        %
+ restricted CTC, $M_1$=5,$M_2$=$40$     & 9.10    & 7.28e-3        \\        %
+ restricted CTC, $M_1$=5,$M_2$=$20$     & 9.14    & 6.93e-3                  %
\end{tabular}
\end{table}

To speed up the recognition by utilizing batch
processing, the order of segments is sorted by their lengths to process
utterances of similar lengths in the same batch.
For fast speech recognition, we gradually performed the batched
beam search in Section~\ref{sect:batch} using a batch size of 21, moved the CTC
prefix score calculation to the CPU, and performed the end-of-speech detection of
Section~\ref{sect:ctc}. As shown in the second, third, and
forth rows of Table~\ref{tab:batch_decoding}, the decoding time is reduced to
543~s, 331~s, and 303~s with considerable accuracy. The decoding
time reduction is obtained owing to accelerating parallelization, the prevention of
sequential processing, and quickly detecting end-of speech. 
\begin{figure}[t!]
	\caption{Cumulative histogram of time index of 
	end-of-speech using baseline and proposed end-of-speech detection 
	methods, respectively.\\}
    \label{fig:eos}
    \centering
		\subfloat[\color{\BLUE}realworld Korean speech]{
				\includegraphics[width=0.50\columnwidth]{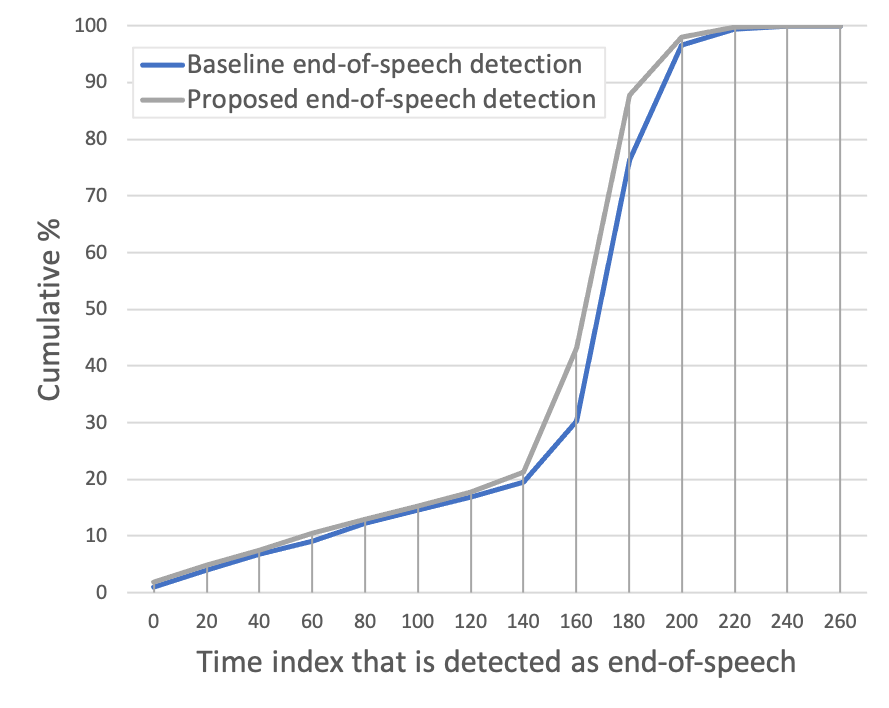}
    }\\
		\subfloat[\color{\BLUE}Librispeech test-clean]{
				\includegraphics[width=0.50\columnwidth]{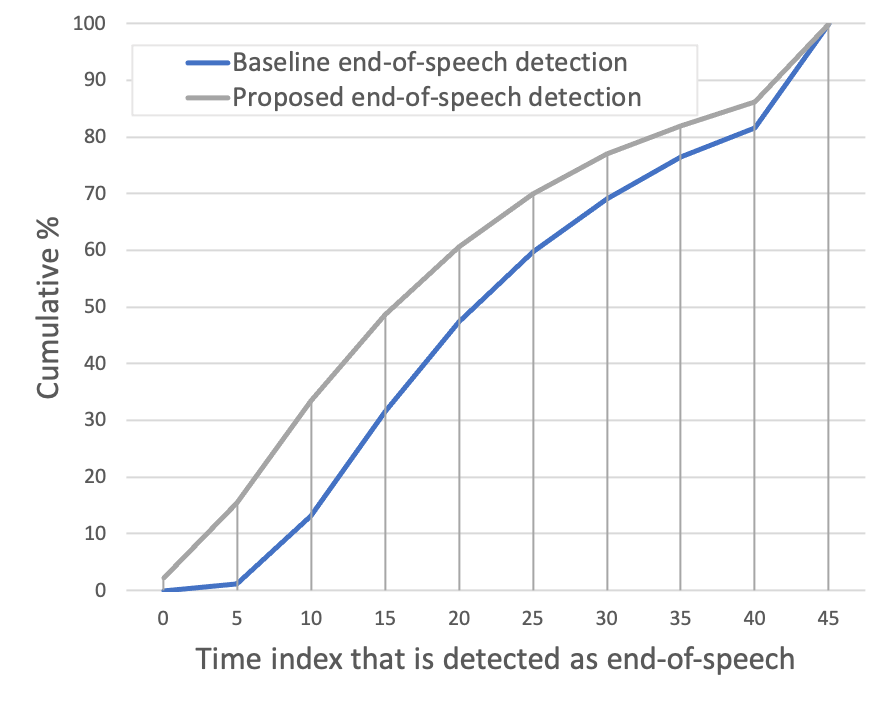}
    }\\
		\subfloat[\color{\BLUE}Librispeech test-other]{
				\includegraphics[width=0.50\columnwidth]{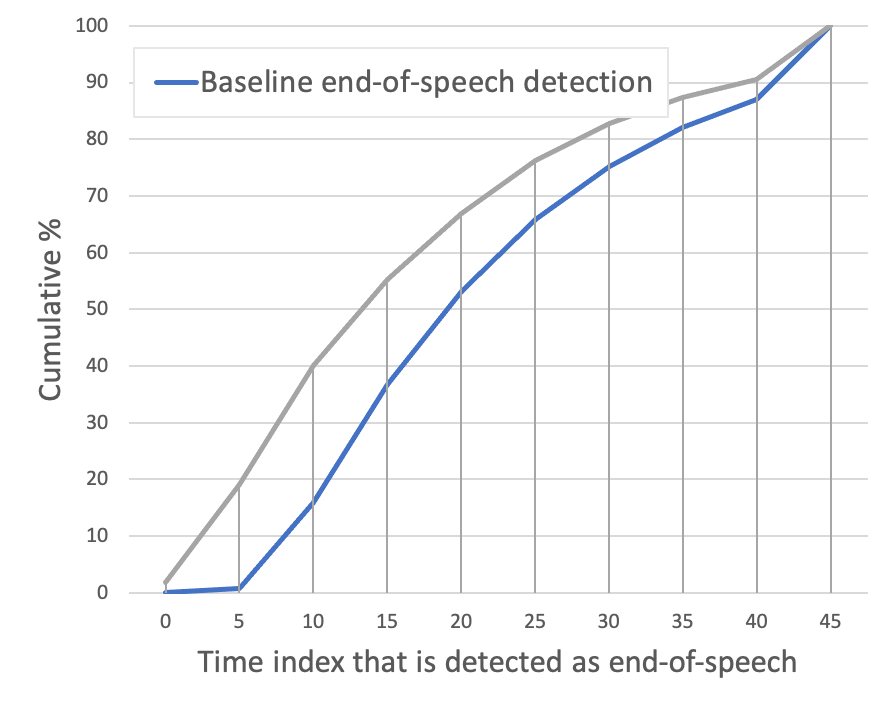}
    }
\end{figure}
To investigate the effect of the proposed end-of-speech detection method, we
calculate the cumulative histogram of the time index of the end-of-speech using the
baseline method and the proposed method, respectively, as shown in
Fig.~\ref{fig:eos}. Note that the values on the $x$-axis are time-index bins of
detected end-of-speech; and the values on the $y$-axis are the normalized accumulated
occurrence for each bin. From the figures, we observe that the proposed method
quickly detects end-of-speech {\color{\BLUE}without loss of the recognition accuracy
compared to the baseline method.}

For a further speedup, the restricted CTC prefix score of Section~
\ref{sect:ctc} can be applied. First, we restricted the start time of the CTC
prefix score with $M_1$ of 5 and then an end time with $M_2$ of
$60$, $40$, or $20$. From the fifth row of Table~\ref{tab:batch_decoding}, the
decoding time was reduced to 263~s with comparable accuracy if only the start time is
restricted. From the last three rows of Table~\ref{tab:batch_decoding},
the decoding time is further reduced to 211~s, 210~s, and 200~s, with an accuracy
degradation according to $M_2$ when the end time is restricted.

%% file: exp_segment.tex
\subsection{Decoding time versus utterance length}

One of the most significant drawbacks of the Transformer model is its weakness in
long sequences because self-attention in each layer computes the output using the
relationship between entire input sequences. It is well-known that the
computation complexity is proportional to the square of the input length~\cite{transformer,reformer}.

In this section, the effect of the sequence length is analyzed using the Librispeech
dataset and the model used in the previous section. The utterances in the Librispeech
dataset are short in length, and the longest utterance is 35~s long. All
utterances in the test set are grouped by lengths as 0--5, 5--10, 10--15,
15--20, and longer than 20~s, and they are decoded separately to measure
the accuracy and recognition speed.  The results are shown in
Table~\ref{tab:uttdur-vs-wer} in two test sets: test-clean and test-other. The two
sets are decoded together to measure the xRTs. The error rates are
computed separately.

As shown in Table~\ref{tab:uttdur-vs-wer}, the error rates are much higher for
the group with the longest utterances. The accuracy is best with
10--20-s-long utterances. The recognition speed in xRT
increases as the lengths are
increased.

\begingroup
\setlength{\tabcolsep}{4pt} %
\begin{table}[t!]
\caption{Dependency of word error rate and recognition speed on
		length of input utterance.
		}
\centering
\begin{tabular}{c|c|cc|cc|c}
\label{tab:uttdur-vs-wer}
    Utterance & Avg. & \multicolumn{2}{c|}{test-clean} & \multicolumn{2}{c|}{test-other} &
    Speed \\ \cline{3-6}
    \raisebox{0.2em}{Len(sec)} & \raisebox{0.2em}{Len(sec)} & \# Utts & WER & \# Utts & WER & In xRT \\ \hline
    0--5   & 3.4  & 1097 & 4.45 & 1422 & 10.09 & 0.024 \\ %
    5--10  & 7.1  & 913  & 3.25 & 1029 & 8.21  & 0.032 \\ %
    10--15 & 12.2 & 373  & 2.96 & 318  & 7.20  & 0.045 \\ %
    15--20 & 17.1 & 148  & 3.19 & 124  & 7.27  & 0.053 \\ %
    20--   & 24.1 & 89   & 8.79 & 46   & 12.25 & 0.069 \\ \hline %
    all    & -    & 2620 & 3.97 & 2939 & 8.60  & 0.040 %
\end{tabular}
\end{table}
\endgroup

\begingroup
\setlength{\tabcolsep}{4pt} %
\begin{table}[t!]
    \caption{\textcolor{\BLUE}{Dependency of word error rate and recognition
    speed on length of input utterance for the concatenated test data.  Original
    utterances are concatenated to constitute longer sentences with maximum
    lengths of 10, 20, 30, 40, and 50 seconds respectively.}}
\centering
    \textcolor{\BLUE}{
\begin{tabular}{c|c|cc|cc|c}
\label{tab:uttdur-vs-wer2}
    Utterance & Avg. & \multicolumn{2}{c|}{test-clean} & \multicolumn{2}{c|}{test-other} &
    Speed \\ \cline{3-6}
    \raisebox{0.2em}{Len(sec)} & \raisebox{0.2em}{Len(sec)} & \# Utts & WER & \# Utts & WER & In xRT \\ \hline
    --10  & 9.2  & 2034  & 3.83 & 2163 & 8.48 & 0.041 \\
    --20 & 16.0 & 1218  & 3.87 & 1192  & 8.37  & 0.054 \\
    --30 & 23.9 & 786  & 12.60 & 763  & 18.29 & 0.075 \\
    --40 & 34.8 & 566  & 30.98 & 545  & 37.05 & 0.101 \\
    --50 & 44.6 & 438  & 46.52 & 428  & 52.52 & 0.135
\end{tabular}}
\end{table}
\endgroup

\begin{figure}[t!]
	\caption{{\color{\BLUE}Dependency of word error rate and recognition speed on length of
		input utterance. The values in Table.~\ref{tab:uttdur-vs-wer2} are shown as
		graphs. Note that the values} on $x$-axis are average lengths, which are
    slightly less than maximum lengths.}
    \label{fig:uttdur-vs-wer}
    \centering
    \subfloat[Average length vs. WER]{
        \includegraphics[width=0.45\columnwidth]{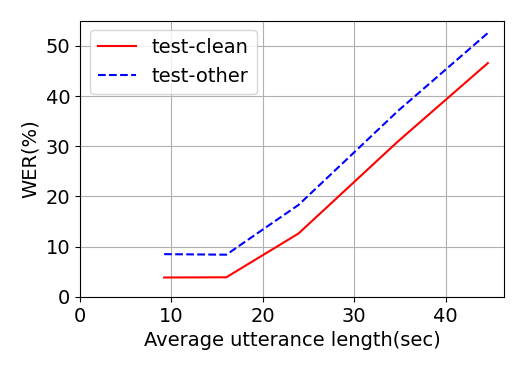}
    }
    \subfloat[Legnth vs. xRT]{
        \includegraphics[width=0.45\columnwidth]{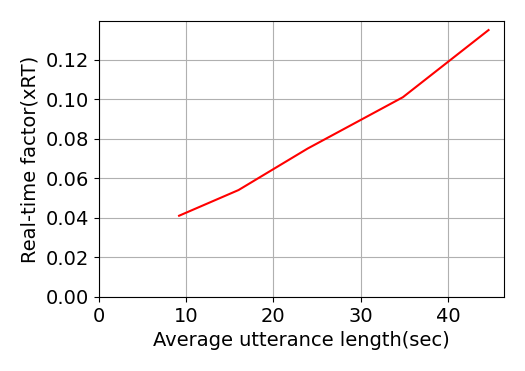}
    }
\end{figure}

To see the effect of the utterance lengths on the recognition performance more
closely, the utterances in the Librispeech test set are concatenated to produce
longer sentences and then fed into the decoder. The reference transcriptions are
also concatenated accordingly.
The concatenation is accomplished as follows: for a given predefined maximum length $L$
and an ordered list of test data, the utterances from the top of the list are
concatenated not to exceed length $L$. If the next utterance makes the concatenated
sentence longer than $L$, then the utterance becomes the beginning of the new sentence, and the next
utterances are concatenated onto this sentence.

As for the maximum length of $L=10,~20,~30,~40~\mathrm{and}~50$, the test sets
are generated, and then WER and xRT are measured in the same way as in
Table~\ref{tab:uttdur-vs-wer}. The results are {\color{\BLUE}presented in Table.~\ref{tab:uttdur-vs-wer2} and
also shown in} Fig.~\ref{fig:uttdur-vs-wer}.
Note that the values on the $x$-axis are the average
length of concatenated sentences and are shorter than the maximum lengths.
It is clear that the longer the sentences, the higher the error rates, and
the longer it takes to be decoded.

In many real-world speech recognition applications, users speak continuously
and spontaneously as in video captioning. Hence it is important to recognize
long utterances quickly and accurately. In a next section, a method to
efficiently decode long utterances is explained.

\subsection{Recognition of long utterance by segmentation}

As presented in the previous section, the input length has great influence on the
performance of the Transformer model.
In Section~\ref{subsec:batch-decoding}, to recognize speech with a Transformer model,
manually segmented utterances are used. However, manual segmentation
cannot be accomplished for large test corpus in real-world applications.
To overcome this issue, we present two automatic segmentation methods in
this section.

First, we applied DNN-based VAD, which split 83 utterances into 8620 short
pieces. These had an average length of 3.34~s and maximum length of 12~s.
Consecutive short pieces are merged into
a segment of longer than minimum length, and pieces longer than the maximum
length are split again uniformly. Second, hard segmentation is applied so that
the lengths of resulting segments are between minimum and maximum length and are
as uniform as possible, including the last segment.
The lengths and numbers of segments for the two methods are given in
Table~\ref{tab:segmentinfo}. For DNN-VAD, min and max lengths are set to 15 and 20~s,
respectively; and for hard segmentation they are set to 19 and 20~s. These
values are selected in consideration of the batch size allowed for the GPU cards
used in the experiments.
The segmented utterances are fed into the end-to-end recognizer corresponding to
the seventh row of Table~\ref{tab:batch_decoding}.
Table~\ref{tab:acc_seg} shows the recognition performance and speed.
{\color{\BLUE}The accuracy with hard segmentation is worse than the
DNN-VAD segmentation as well as the manual segmentation,
since it splits an utterance at any points randomly
even in the middle of a word and words at boundary are recognized incorrectly.
DNN-VAD segmentation method splits utterances in short pauses,
which are usually boundaries of phrases.}

The difference in CERs between manual segmentation and DNN-VAD is mainly owing to
insertion errors since in manual segmentation, overlapped speeches that are difficult
to transcribe are trimmed. The accuracy drop in hard segmentation is owing
to the deletions at segmented boundaries. Further work is ongoing to
reduce this type of error.

{\color{\BLUE}Also the number of segments with DNN-VAD is larger than the manual and the hard
segmentation and the lengths of segments are shorter as shown in
Table~\ref{tab:segmentinfo}.  Thus we could use larger batchsize with DNN-VAD
segmentation and the recognition is faster than with other segmentation methods.
However as pointed out in Section~\ref{sec:longutts}, DNN-VAD requires extra
computation with GPUs to compute DNN outputs, which prevents the method to be
used in commercial use to decode massive data and the hard segmentation method
is preferred over the DNN-VAD.}

\begin{table}[t!]
\caption{Statistics of length and number of segments before and after splitting.}
\centering
\begin{tabular}{c|ccc}
\label{tab:segmentinfo}
Method & \makecell{Num. of\\[-0.5ex] segments} & Avg. Length & Std. Dev \\ \hline
Source & 83 & 347.7 & 464.4 \\ \hline
Manual & 1,173 & 22.8 & 7.28 \\
DNN-VAD & 1,838 & 15.7 & 2.79 \\
Hard Seg. & 1,445 & 20.0 & 1.34
\end{tabular}
\end{table}

\begin{table}[t!]
	\caption{Recognition accuracy in CER (\%) and averaged xRT after three trials.
    Allowed batch size is largest
    applicable in experiment.}
\label{tab:acc_seg}
\centering
\begin{tabular}{c|ccc}
    Method & CER & \makecell{xRT} & \makecell{Allowed\\[-0.5ex]batchsize} \\ \hline
    Manual & 9.10 & 6.98e-3 & 21 \\ %
    DNN-VAD & 10.73 & 5.83e-3 & 66 \\ %
    Hard Seg. & 12.22 & 6.39e-3 & 64 \\ %
\end{tabular}
\end{table}

%% file: conc.tex
\section{Conclusion}
\label{sec:conc}
The Transformer model is being actively used in many research areas yielding a
state-of-the-art performance, and in the field of ASR, the model is leading
the scoreboard. In this work, various methods to bring the model into practical
use were proposed and verified.

This paper proposed fast and efficient recognition methods to
utilize offline Transformer-based end-to-end speech recognition in
real-world applications equipped with low computational resources.
For fast decoding, we adopted a multiple-utterance multiple-hypothesis batched beam search for a
Transformer-based ASR to accelerate a GPU parallelization. The proposed
CTC-based end-of-speech detection could quickly complete the searching process,
thus speeding up the recognition. This
is more effective for noisy and sparsely uttered utterances, where mismatches
between training and testing conditions are significant.
Moreover, the proposed CPU-based and time-restricted CTC prefix score reduced
the computational complexity by limiting the time range to be examined for each
decoding step.
For the efficient decoding
of long speeches, we proposed to split long speeches into segments
using two methods: (a) DNN-VAD-based segmentation and (b) hard segmentation.
The DNN-VAD based segmentation achieved better recognition accuracy than the
hard segmentation. However, the DNN-VAD based segmentation required
a significant amount of computation resources, while hard segmentation could be done without additional
computation. The segmentation of long speeches makes possible stable recognition
of speech from various applications with Transformer models using limited
GPU memories. This ensures stable accuracy and fast speed by boosting the proposed batch processing.

Speech recognition experiments were performed on the Librispeech dataset to
verify the proposed algorithm. Experiments were performed on a
real-world Korean speech dataset recorded at meetings with
multiple speakers. For the 8~h of speech after being segmented,
speech-to-text conversion was undertaken for less than 3~min by a Transformer-based
end-to-end ASR system employing the proposed method with two GPU cards.
Moreover, the ASR system achieved a CER of
10.73\%, which is 27.1\% lower than that of the conventional DNN-HMM based
   ASR system.

%% file: ETRIJ-v2.bbl
\begin{thebibliography}{50}
\providecommand{\natexlab}[1]{#1}
\providecommand{\url}[1]{\texttt{#1}}
\expandafter\ifx\csname urlstyle\endcsname\relax
  \providecommand{\doi}[1]{doi: #1}\else
  \providecommand{\doi}{doi: \begingroup \urlstyle{rm}\Url}\fi

\bibitem[et~al(2019{\natexlab{a}})]{asr_review}
A.~B.~{Nassif} et~al.
\newblock {Speech recognition using deep neural networks: a systematic review}.
\newblock \emph{IEEE Access}, 7:\penalty0 19143--19165, 2019{\natexlab{a}}.

\bibitem[{Roger} et~al.(2020){Roger}, {Farinas}, and {Pinquier}]{asr_review2}
V.~{Roger}, J.~{Farinas}, and J.~{Pinquier}.
\newblock {Deep neural networks for automatic speech processing: a survey from
  large corpora to limited data}.
\newblock In \emph{arXiv preprint, CoRR}, 2020.

\bibitem[et~al(2020{\natexlab{a}})]{asr_review3}
J.~{Li} et~al.
\newblock {On the comparison of popular end-to-end models for large scale
  speech recognition}.
\newblock In \emph{in Proc. Conf. Int. Speech Commun. Assoc.}, pages 1--5,
  Shanghai, China, Dec. 2020{\natexlab{a}}.

\bibitem[{Chung} et~al.(2019){Chung}, {Park}, and {Jung}]{asr_dnn}
H.~{Chung}, J.~G. {Park}, and H.~{Jung}.
\newblock {Rank-weighted reconstruction feature for a robust deep neural
  network-based acoustic model}.
\newblock \emph{ETRI J.}, 41\penalty0 (2):\penalty0 235--241, 2019.

\bibitem[{Sutskever} et~al.(2014){Sutskever}, {Vinyals}, and {Le}]{Sutskever14}
I.~{Sutskever}, O.~{Vinyals}, and O.~V. {Le}.
\newblock {Sequence to sequence learning with neural networks}.
\newblock In \emph{in Proc. Adv. Neural Info. Proc. Sys.}, volume~27, Montreal,
  Canada, Dec. 2014.

\bibitem[{Bahdanau} et~al.(2015){Bahdanau}, {Cho}, and {Bengio}]{Bahdanau2015}
D.~{Bahdanau}, K.~{Cho}, and Y.~{Bengio}.
\newblock {Neural machine translation by jointly learning to align and
  translate}.
\newblock In \emph{in Proc. Int. Conf. Learn. Represent.}, San Diego, CA, USA,
  May 2015.

\bibitem[et~al(2017{\natexlab{a}})]{transformer}
A.~{Vaswani} et~al.
\newblock {Attention is all you need}.
\newblock In \emph{in Proc. Adv. Neural Info. Proc. Sys.}, volume~30, Long
  Beach, CA, USA, Dec. 2017{\natexlab{a}}.

\bibitem[{Dong} et~al.(2018){Dong}, {Xu}, and {Xu}]{speech_transformer}
L.~{Dong}, S.~{Xu}, and B.~{Xu}.
\newblock {Speech-Transformer: a no-recurrence sequence-to-sequence model for
  speech recognition}.
\newblock In \emph{in Proc. IEEE Int. Conf. Acoust., Speech Signal Process.},
  pages 5884--5888, Calgary, Canada, Apr. 2018.

\bibitem[{Moritz} et~al.(2019){Moritz}, {Hori}, and {Roux}]{Moritz19}
N.~{Moritz}, T.~{Hori}, and J.~L. {Roux}.
\newblock {Streaming end-to-end speech recognition with joint CTC-attention
  based models}.
\newblock In \emph{in Proc. IEEE Workshop Auto. Speech Recog. Understanding},
  pages 936--943, Sentosa, Singapore, Dec. 2019.

\bibitem[{Moritz} et~al.(2020){Moritz}, {Hori}, and {Le}]{Moritz20}
N.~{Moritz}, T.~{Hori}, and J.~{Le}.
\newblock {Streaming automatic speech recognition with the Transformer model}.
\newblock In \emph{in Proc. IEEE Int. Conf. Acoust., Speech Signal Process.},
  pages 6074--6078, Barcelona, Spain, Mar. 2020.

\bibitem[{Hwang} and {Lee}(2020)]{transformer_mono}
H.~{Hwang} and C.~{Lee}.
\newblock {Linear-time Korean morphological analysis using an action-based
  local monotonic attention mechanism}.
\newblock \emph{{ETRI J.}}, 42\penalty0 (1):\penalty0 101--107, 2020.

\bibitem[{Parthasarathi} and {Strom}(2019)]{asr_largedata}
S.~H.~K. {Parthasarathi} and N.~{Strom}.
\newblock {Lessons from building acoustic models with a million hours of
  speech}.
\newblock In \emph{in Proc. IEEE Int. Conf. Acoust., Speech Signal Process.},
  pages 6670--6674, Brighton, UK, May 2019.

\bibitem[{Zhou} et~al.(2018){Zhou}, {Xu}, and {Xu}]{transformer_asr_201806}
S.~{Zhou}, S.~{Xu}, and B.~{Xu}.
\newblock {Multilingual end-to-end speech recognition with a single Transformer
  on low-resource languages}.
\newblock In \emph{arXiv preprint, CoRR}, 2018.

\bibitem[et~al(2018{\natexlab{a}})]{transformer_asr_201809}
S.~{Zhou} et~al.
\newblock {Syllable-based sequence-to-sequence speech recognition with the
  Transformer in Mandarin Chinese}.
\newblock In \emph{in Proc. Conf. Int. Speech Commun. Assoc.}, pages 791--795,
  Hyderabad, India, Sep. 2018{\natexlab{a}}.

\bibitem[et~al(2019{\natexlab{b}})]{espnet_transformer2019}
S.~{Karita} et~al.
\newblock {A comparative study on Transformer vs RNN in speech applications}.
\newblock In \emph{in Proc. IEEE Workshop Auto. Speech Recog. Understanding},
  pages 449--456, Sentosa, Singapore, Dec. 2019{\natexlab{b}}.

\bibitem[et~al(2020{\natexlab{b}})]{conformer}
A.~{Gulati} et~al.
\newblock {Conformer: convolution-augmented Transformer for speech
  recognition}.
\newblock In \emph{in Proc. Conf. Int. Speech Commun. Assoc.}, pages
  5036--5040, Shanghai, China, Dec. 2020{\natexlab{b}}.

\bibitem[et~al(2020{\natexlab{c}})]{convtransformer_transducer}
W.~{Huang} et~al.
\newblock {Conv-Transformer Transducer: low latency, low frame rate, streamable
  end-to-end speech recognition}.
\newblock In \emph{in Proc. Conf. Int. Speech Commun. Assoc.}, pages
  5001--5005, Shanghai, China, Dec. 2020{\natexlab{c}}.

\bibitem[et~al(2019{\natexlab{c}})]{joint_ctc_transformer}
S.~{Karita} et~al.
\newblock {Improving Transformer-based end-to-end speech recognition with
  connectionist temporal classification and language model integration}.
\newblock In \emph{in Proc. Conf. Int. Speech Commun. Assoc.}, pages
  1408--1412, Graz, Austria, Sep. 2019{\natexlab{c}}.

\bibitem[et~al(2020{\natexlab{d}})]{online_ctc_att_asr2020}
H.~{Miao} et~al.
\newblock {Transformer-based online CTC/attention end-to-end speech recognition
  architecture}.
\newblock In \emph{in Proc. IEEE Int. Conf. Acoust., Speech Signal Process.},
  pages 6084--6088, Barcelona, Spain, Mar. 2020{\natexlab{d}}.

\bibitem[et~al(2020{\natexlab{e}})]{multispk_asr_w_ctc2020}
X.~{Chang} et~al.
\newblock {End-to-end multi-speaker speech recognition with Transformer}.
\newblock In \emph{in Proc. IEEE Int. Conf. Acoust., Speech Signal Process.},
  pages 6134--6138, Barcelona, Spain, Mar. 2020{\natexlab{e}}.

\bibitem[et~al(2020{\natexlab{f}})]{nonar_w_ctc2020}
Y.~{Higuchi} et~al.
\newblock Mask {CTC:} non-autoregressive end-to-end {ASR} with {CTC} and mask
  predict.
\newblock In \emph{in Proc. Conf. Int. Speech Commun. Assoc.}, pages
  3655--3659, Shanghai, China, Dec. 2020{\natexlab{f}}.

\bibitem[et~al(2020{\natexlab{g}})]{insertion_based_modeling_w_ctc2020}
Y.~{Fujita} et~al.
\newblock {Insertion-based modeling for end-to-end automatic speech
  recognition}.
\newblock In \emph{in Proc. Conf. Int. Speech Commun. Assoc.}, pages
  3660--3664, Shanghai, China, Dec. 2020{\natexlab{g}}.

\bibitem[{Parcollet} et~al.(2020){Parcollet}, {Morchid}, and
  {Linarès}]{e2e_sincnet2020}
T.~{Parcollet}, M.~{Morchid}, and G.~{Linarès}.
\newblock {E2E-SINCNET: toward fully end-to-end speech recognition}.
\newblock In \emph{in Proc. IEEE Int. Conf. Acoust., Speech Signal Process.},
  pages 7714--7718, Barcelona, Spain, Mar. 2020.

\bibitem[et~al(2020{\natexlab{h}})]{self_mixed_att_2020}
X.~{Zhou} et~al.
\newblock {Self-and-mixed attention decoder with deep acoustic structure for
  Transformer-based LVCSR}.
\newblock In \emph{in Proc. Conf. Int. Speech Commun. Assoc.}, pages
  5016--5020, Shanghai, China, Dec. 2020{\natexlab{h}}.

\bibitem[et~al(2020{\natexlab{i}})]{nonar_w_ctc2020v2}
Y.~{Higuchi} et~al.
\newblock {Improved mask-CTC for non-autoregressive end-to-end ASR}.
\newblock In \emph{arXiv preprint, CoRR}, 2020{\natexlab{i}}.

\bibitem[et~al(2020{\natexlab{j}})]{self_dist_2020}
T.~{Moriya} et~al.
\newblock {Self-distillation for improving CTC-Transformer-based ASR systems}.
\newblock In \emph{in Proc. Conf. Int. Speech Commun. Assoc.}, pages 546--550,
  Shanghai, China, Dec. 2020{\natexlab{j}}.

\bibitem[et~al(2020{\natexlab{k}})]{long_context_2020}
T.~{Hori} et~al.
\newblock {Transformer-based long-context end-to-end speech recognition}.
\newblock In \emph{in Proc. Conf. Int. Speech Commun. Assoc.}, pages
  5011--5015, Shanghai, China, Dec. 2020{\natexlab{k}}.

\bibitem[et~al(2020{\natexlab{l}})]{cross_att_2020}
Y.~{Zhao} et~al.
\newblock {Cross attention with monotonic alignment for speech Transformer}.
\newblock In \emph{in Proc. Conf. Int. Speech Commun. Assoc.}, pages
  5031--5035, Shanghai, China, Dec. 2020{\natexlab{l}}.

\bibitem[et~al(2020{\natexlab{m}})]{code_switching_2020}
Y.~{Lu} et~al.
\newblock {Bi-encoder Transformer network for Mandarin-English code-switching
  speech recognition using mixture of experts}.
\newblock In \emph{in Proc. Conf. Int. Speech Commun. Assoc.}, pages
  4766--4770, Shanghai, China, Dec. 2020{\natexlab{m}}.

\bibitem[et~al(2020{\natexlab{n}})]{adapt_adjust_2020}
G.~I.~{Winata} et~al.
\newblock {Adapt-and-adjust: overcoming the long-tail problem of multilingual
  speech recognition}.
\newblock In \emph{arXiv preprint, CoRR}, 2020{\natexlab{n}}.

\bibitem[et~al(2020{\natexlab{o}})]{lightweight_2020}
L.~{Kürzinger} et~al.
\newblock {Lightweight end-to-end speech recognition from raw audio data using
  Sinc-convolutions}.
\newblock In \emph{in Proc. Conf. Int. Speech Commun. Assoc.}, pages
  1659--1663, Shanghai, China, Dec. 2020{\natexlab{o}}.

\bibitem[et~al(2020{\natexlab{p}})]{unsup_pretraining_mtkl_2020}
S.~{Li} et~al.
\newblock {Improving Transformer-based speech recognition with unsupervised
  pre-training and multi-task semantic knowledge learning}.
\newblock In \emph{in Proc. Conf. Int. Speech Commun. Assoc.}, pages
  5006--5010, Shanghai, China, Dec. 2020{\natexlab{p}}.

\bibitem[et~al(2016)]{deepspeech2}
D.~{Amodei} et~al.
\newblock {Deep Speech 2 : end-to-end speech recognition in English and
  Mandarin}.
\newblock In \emph{in Proc. Int. Conf. Machine Learning}, pages 173--182, New
  York, NY, USA, June 2016.

\bibitem[et~al(2020{\natexlab{q}})]{nvidiabatch}
H.~{Braun} et~al.
\newblock {GPU-accelerated Viterbi exact lattice decoder for batched online and
  offline speech recognition}.
\newblock In \emph{in Proc. IEEE Int. Conf. Acoust., Speech Signal Process.},
  pages 7874--7878, Barcelona, Spain, Mar. 2020{\natexlab{q}}.

\bibitem[{Oh} et~al.(2020){Oh}, {Park}, and {Park}]{batchBLSTM}
Y.~R. {Oh}, K.~{Park}, and J.G. {Park}.
\newblock {Online speech recognition using multichannel parallel acoustic score
  computation and deep neural network (DNN)-based voice-activity detector}.
\newblock \emph{Appl. Sci.}, 10\penalty0 (12), 2020.

\bibitem[{Seki} et~al.(2018){Seki}, {Hori}, and
  {Watanabe}]{e2e_vector_beam_search2018}
H.~{Seki}, T.~{Hori}, and S.~{Watanabe}.
\newblock {Vectorization of hypotheses and speech for faster beam search in
  encoder decoder-based speech recognition}.
\newblock In \emph{arXiv preprint, CoRR}, 2018.
\newblock URL \url{arXiv:cs/1811.04568}.

\bibitem[et~al(2019{\natexlab{d}})]{e2e_vector_beam_search}
H.~{Seki} et~al.
\newblock {Vectorized beam search for CTC-attention-based speech recognition}.
\newblock In \emph{in Proc. Conf. Int. Speech Commun. Assoc.}, pages
  3825--3829, Graz, Austria, Sep. 2019{\natexlab{d}}.

\bibitem[et~al(2006)]{ctc}
A.~{Graves} et~al.
\newblock {Connectionist temporal classification: labelling unsegmented
  sequence data with recurrent neural networks}.
\newblock In \emph{in Proc. Int. Conf. Machine Learning}, page 369–376,
  Pittsburgh, PA, USA, June 2006.

\bibitem[et~al(2020{\natexlab{r}})]{restricted_ctc}
H.~{Miao} et~al.
\newblock {Online hybrid CTC/attention end-to-end automatic speech recognition
  architecture}.
\newblock \emph{IEEE TASLP}, 28:\penalty0 1452--1465, 2020{\natexlab{r}}.

\bibitem[et~al(2020{\natexlab{s}})]{Yoshimura20}
T.~{Yoshimura} et~al.
\newblock {End-to-end automatic speech recognition integrated with CTC-based
  voice activity detection}.
\newblock In \emph{in Proc. IEEE Int. Conf. Acoust., Speech Signal Process.},
  pages 6999--7003, Barcelona, Spain, Mar. 2020{\natexlab{s}}.

\bibitem[{Hori} et~al.(2017){Hori}, {Watanabe}, and {Hershey}]{joint_ctc_att}
T.~{Hori}, S.~{Watanabe}, and J.~{Hershey}.
\newblock {Joint {CTC}/attention decoding for end-to-end speech recognition}.
\newblock In \emph{in Proc. Ann. Mtg. Assoc. Comp. Ling.}, pages 518--529,
  Vancouver, Canada, July 2017.

\bibitem[{Meister} et~al.(2020){Meister}, {Vieira}, and
  {Cotterell}]{bestfirst_beamsearch}
C.~{Meister}, T.~{Vieira}, and R.~{Cotterell}.
\newblock {Best-first beam search}.
\newblock \emph{TACL}, 8:\penalty0 795--809, 2020.

\bibitem[et~al(2017{\natexlab{b}})]{hybrid_ctc_att}
S.~{Watanabe} et~al.
\newblock {Hybrid CTC/attention architecture for end-to-end speech
  recognition}.
\newblock \emph{IEEE J-STSP}, 11\penalty0 (8):\penalty0 1240--1253,
  2017{\natexlab{b}}.

\bibitem[et~al(2019{\natexlab{e}})]{relativePE_transformer}
P.~{Zhou} et~al.
\newblock {Improving generalization of Transformer for speech recognition with
  parallel schedule sampling and relative positional embedding}.
\newblock In \emph{arXiv preprint, CoRR}, volume abs/1911.00203,
  2019{\natexlab{e}}.

\bibitem[{Kitaev} et~al.(2020){Kitaev}, {Kaiser}, and {Levskaya}]{reformer}
N.~{Kitaev}, L.~{Kaiser}, and A.~{Levskaya}.
\newblock {Reformer: the efficient Transformer}.
\newblock In \emph{in Proc. Int. Conf. Learn. Represent.}, Addis Ababa,
  Ethiopia, Apr. 2020.

\bibitem[{Park}(2013)]{epd}
K.~{Park}.
\newblock {A robust endpoint detection algorithm for the speech recognition in
  noisy environments}.
\newblock In \emph{{in Proc. INTER-NOISE and NOISE-CON Congress and Conf.}},
  volume 247, pages 5790--5795, Innsbruck, Austria, Sep. 2013.

\bibitem[et~al(2018{\natexlab{b}})]{espnet}
S.~{Watanabe} et~al.
\newblock {ESPnet: end-to-end speech processing toolkit}.
\newblock In \emph{in Proc. Conf. Int. Speech Commun. Assoc.}, pages
  2207--2211, Hyderabad, India, Sep. 2018{\natexlab{b}}.

\bibitem[et~al(2019{\natexlab{f}})]{transformer_cache}
T.~{Xiao} et~al.
\newblock {Sharing attention weights for fast Transformer}.
\newblock In \emph{in Proc. Int. Jnt. Conf. AI}, pages 5292--5298, Macao,
  China, Aug. 2019{\natexlab{f}}.

\bibitem[et~al(2018{\natexlab{c}})]{accum_grad}
M.~{Ott} et~al.
\newblock {Scaling neural machine translation}.
\newblock In \emph{in Proc. Conf. Machine Translation}, pages 1--9, Brussels,
  Belgium, Oct. 2018{\natexlab{c}}.

\bibitem[et~al(2020{\natexlab{t}})]{Bang2020}
J.~{Bang} et~al.
\newblock {Automatic construction of a large-scale speech recognition database
  using multi-genre broadcast data with inaccurate subtitle timestamps}.
\newblock \emph{IEICE Trans Inf Syst}, E103.D\penalty0 (2):\penalty0 406--415,
  2020{\natexlab{t}}.

\end{thebibliography}
